\documentclass[aps,twocolumn,prl,superscriptaddress,preprintnumbers,showpacs,tightenlines]{revtex4}
\usepackage{amssymb}
\usepackage{epsfig,graphicx,times}

\newcommand{\beq}{\begin{equation}}
\newcommand{\eeq}{\end{equation}}
\newcommand{\bea}{\begin{eqnarray}}
\newcommand{\eea}{\end{eqnarray}}

\begin{document}

\preprint{First Draft}
\title{Universality in Exact Quantum State Population Dynamics and Control}
\author{Lian-Ao Wu}
\affiliation{IKERBASQUE, Basque Foundation for Science, 48011, Bilbao, Spain}
\affiliation{Department of Theoretical Physics and History of Science, The Basque Country
University (EHU/UPV), PO Box 644, 48080 Bilbao, Spain}
\author{Dvira Segal}
\affiliation{Department of Chemistry and Center for Quantum Information and Quantum
Control, University of Toronto, 80 St. George Street, Toronto, Ontario M5S
3H6, Canada}
\author{\'I\~nigo L. Egusquiza}
\affiliation{Department of Theoretical Physics and History of Science, The Basque Country
University (EHU/UPV), PO Box 644, 48080 Bilbao, Spain}
\author{Paul Brumer}
\affiliation{Department of Chemistry and Center for Quantum Information and Quantum
Control, University of Toronto, 80 St. George Street, Toronto, Ontario M5S
3H6, Canada}
\date{\today }

\begin{abstract}
We consider an exact population transition, defined as the probability of
finding a state at a final time  being exactly equal to the probability
of another state
at the initial time. We prove that, given a Hamiltonian, there always
exists a complete set of orthogonal states that can be employed as
time-zero states for which this exact population transition occurs.
The result is general: it holds for arbitrary systems, arbitrary
pairs of initial and final states, and for any time interval. The
proposition is illustrated with several analytic models. In particular
we demonstrate that in some cases, by tuning the control parameters
a \textit{complete} transition might occur, where a target state,
vacant at $t=0$, is fully populated at time $\tau$.
\end{abstract}

\maketitle


\emph{Introduction.---} The central goal of quantum control is the
transfer of population from an initial state to a final target state
\cite{Bookrice,BSbook}. Within the framework of coherent quantum
control, focus has been primarily on designing specific laser-based
scenarios that achieve this goal (for some bound state examples see,
e.g., \cite{shapiro,jiangbin,Kbergmann}), whereas within the
framework of optimal control, focus has been on identifying control
fields that achieve this goal, both computationally and
experimentally.

Despite the enormous interest in this area there are very few
analytic control results about realistic systems. These include
theorems such as that of Huang-Tarn-Clark \cite{HTC1983}, a theorem
by Ramakrisnan $et$ $al$. on the dimensionality of the Lie Algebra
induced by the interaction between the system and the control field
\cite{Rama1995}, and a theorem by Shapiro and Brumer
\cite{Brumer95}, where control was shown to depend on the
dimensionality of the controlled subspaces. As a consequence, any
proven fundamental result adds considerably to the knowledge base
(e.g., \cite{Openwu}). In this paper we expose a universal feature
of quantum {\it dynamics} that has significant implications for
control. The focus here is in the proving this dynamical result;
future studies will be directed to control applications.

Specifically, consider an initial state $|\Psi(0)\rangle$ that
evolves under a Hamiltonian $H$ to yield the state $|\Psi(\tau)
\rangle$ at time $t=\tau$. Of interest is the probability $P_I(0) =
|\langle I|\Psi(0)\rangle|^2$ of the system being initially in state
$|I\rangle$ undergoing a transition with probability $P_F(\tau) =
|\langle F|\Psi(\tau)\rangle|^2$ to an orthogonal component
$|F\rangle$ at time $\tau$. We focus on the possibility of an
``exact quantum transition" between these states defined as
\begin{eqnarray}
P_{F}(\tau )=P_{I}(0),  \label{eq:est}
\end{eqnarray}
i.e., where the probability of observing state $|F\rangle$ at final
time $\tau$ equals the probability of observing the state
$|I\rangle$ initially.

In this paper we prove that there always exists,
for arbitrary evolution operator and for an arbitrary time $\tau$,
a complete set $\left\{ \Psi
_{k}(0)\right\}$ of orthogonal states that undergo the exact
state transition (\ref{eq:est}) from $|I\rangle$ to $|F\rangle$.
For a given Hamiltonian, the magnitude of the associated $P_F(\tau)$
is determined by the choice of $\tau$, $|I\rangle$ and $|F\rangle$.
As examples, we obtain the set $\left\{\Psi _{k}(0)\right\}$ for
some analytical models, and furthermore provide instances of
\textit{significant transfer}, defined by $P_F(0)\ll P_I(0)$.

While this universality might seem surprising, we show below that it
simply stems from unitarity of quantum evolution. Based on unitary
evolution,  the universal existence of exact quantum state
transmission between different subspaces was demonstrated in
\cite{Wu09},  and cyclic quantum evolution in the theory of
geometric phase was established in \cite{Wu94}. Unitarity is also at
the heart of the no-cloning theorem, which is fundamental to quantum
information science. In the present case an inclusive theorem in
quantum dynamics based on unitarity is derived that is  expected to
be influential in quantum technologies. In particular, note that the
dynamical principle is here established within the {\it same}
Hilbert space, unlike \cite{Wu09}, giving an approach that  is
propitious for a broad range of applications, e.g., for quantum
computing \cite{Chuang}, coherent control of atomic and molecular
processes \cite{BSbook},  and laser control of chemical reaction in
molecules \cite{Bookrice}.

\textit{Universality of the exact population transition.---}
Consider an $M$ dimensional system ($M$ can be infinite), spanned by the
bases $\left\{ \left\vert \alpha \right\rangle \right\}$
and described by density matrix $\rho$.
The equality in Eq. (\ref{eq:est}) becomes
\begin{eqnarray}
\text{tr}[\left\vert F\right\rangle \left\langle F\right\vert \rho (\tau )]=
\text{tr}[\left\vert I\right\rangle \left\langle I\right\vert \rho (0)].
\label{eq:exact}
\end{eqnarray}
We assume that the system is prepared in a pure state $\Psi(0)$ so that
$\rho(0)= |\Psi (0)\rangle \langle \Psi(0)|$.

\textbf{Proposition.} There always exists a complete orthogonal set $\left\{
\Psi _{k}(0)\right\} _{\tau }$, which depends on $\tau$, such that an exact
population transition described by Eqs.~(\ref{eq:est}) or (\ref{eq:exact})
takes place if the initially prepared state is a member of this set.

\textbf{Proof.} Assuming that at time $t=0$ the state of the system is $\Psi
(0)$, the left side of Eq. (\ref{eq:exact}) can be written as
\begin{eqnarray}
\text{tr}\left[ \left\vert F\right\rangle \left\langle F\right\vert \rho
(\tau )\right] &=&\text{tr}[\left\vert F\right\rangle \left\langle
F\right\vert U(\tau )\rho (0)U^{\dagger }(\tau )]  \nonumber \\
&=&\left\langle \Psi (0)\right\vert U^{\dagger }(\tau )\left\vert
F\right\rangle \left\langle F\right\vert U(\tau )\left\vert \Psi
(0)\right\rangle  \nonumber \\
&=&\left\langle \Psi (0)\right\vert U^{\dagger }(\tau )\mathcal{E}\left\vert
I\right\rangle \left\langle I\right\vert \mathcal{E}U(\tau )\left\vert \Psi
(0)\right\rangle  \nonumber \\
&=&\text{tr}[\left\vert I\right\rangle \left\langle I\right\vert \rho
^{\prime }(\tau )],  \label{eq:b}
\end{eqnarray}
where $U(\tau)$ is the time evolution operator of the system, and we have
introduced the exchange operator
\bea
\mathcal{E}=\left\vert F\right\rangle \left\langle I\right\vert +\left\vert
I\right\rangle \left\langle F\right\vert +\mathcal{E}_{0},
\eea
satisfying $\mathcal{E}\left\vert I\right\rangle \left\langle I\right\vert
\mathcal{E}=\left\vert F\right\rangle \left\langle F\right\vert $, with $%
\mathcal{E}_{0}=\sum_{\alpha \neq I,F}^{M}\left\vert \alpha \right\rangle
\left\langle \alpha \right\vert $. The exchange operator $\mathcal{E}$ swaps
the states $\left\vert F\right\rangle $ and $\left\vert I\right\rangle$
while keeping other states intact. It is easy to prove that
$\mathcal{E}^{2}=1$. We have also defined the auxiliary density matrix $\rho ^{\prime
}(\tau )=\left\vert \Psi ^{\prime }(\tau )\right\rangle \left\langle \Psi
^{\prime }(\tau )\right\vert $, with $\Psi ^{\prime }(\tau )=W(\tau )\Psi
(0)$ and $W(\tau)=\mathcal{E}U(\tau )$. This operator behaves similarly
to the time-evolution operator. It is significant to
note that the operator $W(\tau )$ is unitary, satisfying
\bea
W^{\dagger }(\tau )W(\tau )=U^{\dagger }(\tau )\mathcal{EE}U(\tau )=1.
\label{eq:unit}
\eea
As a unitary operator $W(\tau)$ can be diagonalized to yield a complete
set of orthonormal eigenvectors $\left\{\Psi_{k}(0)\right\} _{\tau }$ and
exponential eigenvalues $\left\{ \exp (i\phi _{k})\right\} _{\tau }$.
Any vector $\Psi _{k}(0)$ in the set thus obeys the eigenequation
\bea
W(\tau )\Psi _{k}(0)=\exp (i\phi _{k})\Psi _{k}(0).
\label{eq:eigen}
\eea
Comparing Eq.~(\ref{eq:exact}) with Eq.~(\ref{eq:b}), we note that
if the state of the system at time zero $\Psi (0)$ is one of the
$\Psi _{k}(0)$'s in Eq. (\ref{eq:eigen}), then the equality tr$[\left\vert
I\right\rangle \left\langle I\right\vert \rho (0)]=$tr$[\left\vert
F\right\rangle \left\langle F\right\vert \rho (\tau )]$ in
Eq.~(\ref{eq:exact}) holds. In other words, an exact population
transition occurs between states $|I\rangle$ and $|F\rangle$ independent
of the choice of these states other than that they are orthogonal.
The result is also valid for the exchange operator
%
$\mathcal{E}=e^{i\alpha }\left\vert F\right\rangle \left\langle
I\right\vert
+e^{i\beta }\left\vert I\right\rangle \left\langle F\right\vert +\mathcal{E}%
_{0}$,
%
where $\alpha $ and $\beta$\ are real numbers. In this case the unitary
condition (\ref{eq:unit}) translates to
%
$W^{\dagger }(\tau )W(\tau )=U^{\dagger }(\tau )\mathcal{E}^{\dagger }%
\mathcal{E}U(\tau )=1$,
%
since $\mathcal{E}^{\dagger }$ may not be equal to $\mathcal{E}$.

The above result is a fundamental attribute of quantum dynamics and
should serve as a basic building block in quantum control theory
(see also \cite{Wu09} and \cite{Wu94}). Given an arbitrary
Hamiltonian at an arbitrary time $\tau$, and an arbitrary pair of
states $\left\vert I\right\rangle$ and $\left\vert F\right\rangle$,
$W(\tau)$ can be numerically diagonalized to obtain its eigenvalue
spectrum and eigenstates, giving the states $|I\rangle$ and
$|F\rangle$ between which the "exact transition" $P_I(0)=P_F(\tau)$
takes place. This can be readily done for small systems, and we
illustrate below several simple eigenproblems where the spectrum of
$W(\tau)$ can be analytically obtained. However, we emphasize that,
unlike specific control scenarios, this result is {\it universal},
arising only from the fact that a unitary operator possesses a
complete set of orthogonal eigenvectors. Of particular interest in
control scenarios are transitions, which we denote as {\it
significant}, when $P_I(0)>P_F(0)$. Ideally, in control scenarios,
we seek exact transitions that are (what we term) {\it complete},
i.e. where $P_I(0)=1$ and $P_F(0)=0$, so that an initial state,
fully populated at time zero, transfers its population to a target
state at time $\tau$.
Experience gained from individual sample cases below sheds light on
the theorem and will allow one to assess future directions for
control applications.


\emph{Example: A two level system.---} We discuss three variants of
the two-level-system (TLS) model. In the first \textit{static} case
the theorem holds in a trivial way, though there is no actual
population transition. In the second case a weak time dependent
perturbation leads to a significant population transfer. The last
example demonstrates that in a delta-kicked TLS a complete
transition might take place. The unperturbed TLS model is described
by the states $|0\rangle$ and $|1 \rangle$ of energies $E_0$ and
$E_1$ respectively. Taking into account  different types of
interactions, we explore next the transition between
$|I\rangle=|0\rangle$ to $|F\rangle=|1\rangle$.

\emph{I. Static two-level system.---} Assuming for simplicity that
$E_1=0$, we obtain the eigenstates of $W(\tau)$,
 $\Psi_{\pm}(0)=\frac{1}{\sqrt{2}}(\left\vert 0\right\rangle \pm \exp
(-iE_0 \tau /2)\left\vert 1\right\rangle )$ with eigenvalues
$e^{i\phi_{\pm}}$; $\phi_{+}=-E_0\tau /2$ and $\phi_{-}=- E_0\tau
/2+\pi$. This case is trivial since there is no actual transition
during the course of time. However, the dynamics is still
Hamiltonian and hence (\ref{eq:exact}) is valid.

\begin{figure}[tbp]
\vspace{3mm} {\hbox{\epsfxsize=75mm \epsffile{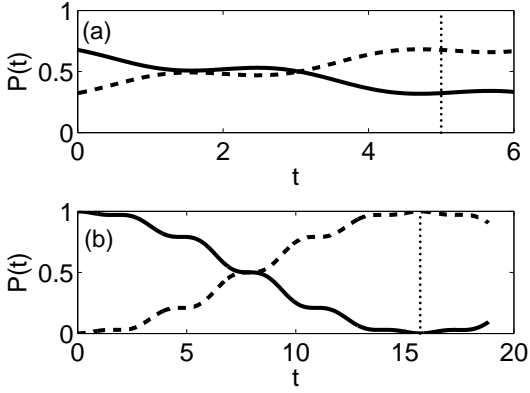}}}
\caption{The dynamics of a two-level system under a cosine
perturbation. $E_1-E_0=\protect\omega=1$, $\lambda=0.2$, $\langle 0|
V|1\rangle = \langle 1|V|0\rangle =1$. (a) An exact transition with
$\protect\tau=5$; (b) A complete transition with
$\protect\tau=\protect\pi/\lambda$; $P_I(t)$ (full line), $P_F(t)$
(dashed line). The dotted line marks the position of $\protect\tau$,
where an exact transition takes place, $P_I(0)=P_F(\protect\tau)$. }
\label{Fig1}
\end{figure}

\emph{II. Two-level system under a time dependent perturbation.---}
Consider next the two states $\left\vert 0\right\rangle $ and
$\left\vert 1\right\rangle $ of energies $E_{0}$ and $E_{1}$
respectively, on-resonance with a periodic perturbation
$H^{\prime}(t)=\lambda V\cos \omega t$, where $\lambda$ is a
parameter characterizing the order of the perturbation expansion.
Setting again 
$\left\vert I\right\rangle =\left\vert 0\right\rangle $, $\left\vert
F\right\rangle =\left\vert 1\right\rangle $, we obtain the
approximate eigenstates of $W(\tau)$ to first order of $\lambda$,
\begin{eqnarray}
\Psi_{\pm }(0) &\approx &\frac{1}{\sqrt{2}(1\pm \frac{r}{2}\cos \theta )}
\left[ -(r\cos \theta \pm 1)\left\vert 0\right\rangle +\left\vert
1\right\rangle \right] ;  \nonumber \\
e^{i\phi _{\pm}} &\approx &\mp e^{\pm ir\sin \theta}
\label{eq:pert1}
\end{eqnarray}
where $re^{i\theta}=i\lambda \int_{0}^{\tau}ds
e^{i(E_{0}-E_{1})s} \cos(\omega s)
\left\langle 1\right\vert V\left\vert 0\right\rangle$; $r$ and $\theta$ are  real numbers.
If $\cos\theta >0$ we obtain
the following inequality for the initial state $\Psi _{+}(0)$
\begin{eqnarray}
\left\vert \left\langle I|\Psi _{+}(0)\right\rangle \right\vert &\approx &
\frac{1}{\sqrt{2}}(1+\frac{r}{2}\cos \theta )  \nonumber \\
&>&\left\vert \left\langle F|\Psi_{+}(0)\right\rangle \right\vert \approx
\frac{1}{\sqrt{2}}(1-\frac{r}{2}\cos \theta ).
\end{eqnarray}
Since $\left\vert \left\langle I|\Psi _{+}(0)\right\rangle \right\vert
> \left\vert \left\langle F|\Psi _{+}(0)\right\rangle \right\vert$, a
significant transfer is realized here.
For $\cos \theta <0$ the same inequality holds for $\Psi _{-}(0)$.
In both cases the initial and final probabilities satisfy
\bea P_{I}(0)=P_{F}(\tau )\approx \frac{1}{2}(1+r|\cos \theta |).
\label{eq:pert2}
\eea
Figure \ref{Fig1} demonstrates an exact population transfer in
the present model, computed without approximation for the
evolution of the TLS under a harmonic perturbation.
Panel (a) demonstrates a significant transition, while for a different set of parameters
panel (b) shows that at specific times (or for a designed time dependent field) a complete
transition might take place, even for weak perturbations.

The analytic calculation (\ref{eq:pert1})-(\ref{eq:pert2}) exemplifies a ``significant
transition", i.e. where
$\left\vert \left\langle I|\Psi _{k}(0)\right\rangle \right\vert > \left\vert
\left\langle F|\Psi _{k}(0)\right\rangle \right\vert$.
Having demonstrated that population transfer can be achieved, we
further address the question, particularly relevant to control, of
what is the maximum achievable significance, defined as
$P_{F}(0)-P_{I}(0)$. For the TLS case the significance equals
$(-\rm{Tr}[\sigma_z\rho])$, which we want to maximize under the
conditions that (a) the state is pure, and (b) an exact quantum
transition is achieved at time $\tau$. It can be shown that the
condition for exact transition can be rewritten as
$r_3=-\vec{s}\cdot\vec{r}$, where we define the vectors
$\vec{s}(\tau)=\frac{1}{2}\mathrm{Tr}\left[
 U^{\dagger }(\tau )\sigma _{z}U(\tau )\vec{\sigma}\right]$
 and $\vec{r}=\rm{Tr}[\rho\vec\sigma]$;  $\sigma_{i=1,2,3}$
are the $x,y,z$ Pauli matrices, respectively.
If, for example, the component $s_3$ was equal to 1, the condition
for an exact transition forces $r_3$ to be zero, and there is no
initial state for which there is significant exact transfer.
Assuming that $s_3\neq1$, on the other hand, the significance
becomes maximal for the initial state \bea \Psi(0) =
\frac{1}{\sqrt{2(1-s_3)}}\left[ (s_{1}-is_{2})|0\rangle +\left(1-
s_{3}\right) |1\rangle \right] \,. \eea
Note then that generally, states that maximize the significance will
{\it not} be eigenstates of $W(\tau)$. Thus, although in accord with
the above theorem one achieves exact transitions, the ideal complete
transition is obtained rarely. However, the proven theorem provides
a new framework in within which to modify the Hamiltonian to achieve
transitions with increasingly larger significance.

\begin{figure}[tbp]
\vspace{3mm} {\hbox{\epsfxsize=80mm \epsffile{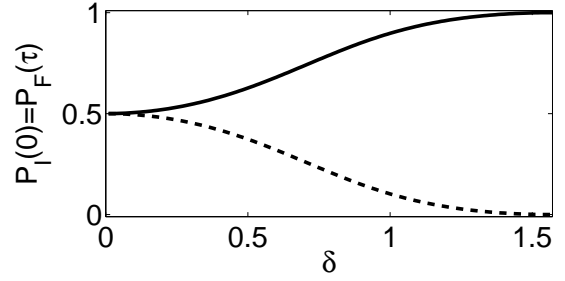}}}
\caption{The population $P_I(0)=P_F(\tau)$  for a kicked two-level
model, showing that with proper tuning of parameters a complete
population transition can  be achieved. $\tau=\pi$,
$\epsilon=\pi/2$, $\omega=1$, $\varpi=1$. $|I\rangle=|0\rangle$
(full line); $|I\rangle=|1\rangle$ (dashed line), } \label{Fig2}
\end{figure}

\emph{III. Kicked two-level system.---}
For the same model, again with $\left\vert I\right\rangle =\left\vert
0\right\rangle $ and $\left\vert F\right\rangle =\left\vert 1\right\rangle$,
consider a \textit{non-perturbative} time-dependent Hamiltonian,
\bea
H=\left\{
\begin{array}{c}
\varpi (\sigma_{z}\cos \epsilon +\sigma _{x}\sin \epsilon ),0<t<\delta \\
\omega \sigma_{z},t>\delta
\end{array}
\right. ,
\eea
where $\sigma_{x}$,  $\sigma_{y}$ and $\sigma_{z}$ are the Pauli
matrices. Here one can generically write the operator
$W(\tau)=-i(\cos \theta +i\overrightarrow{n}\cdot
\overrightarrow{\sigma }\sin \theta)$, where in this case the
parameters are
\begin{eqnarray}
\cos \theta &=&\sin (\varpi \delta )\sin (\epsilon )\cos [\omega (\tau
-\delta )],  \nonumber \\
n_{z}\sin \theta &=&\sin (\varpi \delta )\sin (\epsilon )\sin [\omega (\tau
-\delta )],  \nonumber \\
n_{y}\sin \theta &=&-\cos (\varpi \delta )\sin [\omega (\tau -\delta )]
\nonumber \\
&&-\cos (\epsilon )\cos [\omega (\tau -\delta )]\sin (\varpi \delta),
\nonumber \\
n_{x}\sin \theta &=&-\cos (\epsilon )\sin (\varpi \delta )\sin [\omega (\tau
-\delta )],  \nonumber \\
&&+\cos (\varpi \delta )\cos [\omega (\tau -\delta )],
\end{eqnarray}
with $n_x^2+n_y^2+n_z^2=1$.
The two eigenstates of $W(\tau)$ are $\Psi_{+}(0)=\sqrt{\frac{1+n_{z}}{2}}
(-i)e^{i\gamma}\left\vert 0\right\rangle +\sqrt{\frac{1-n_{z}}{2}}\left\vert
1\right\rangle $ with eigenvalue $-ie^{i\theta }$ and $\Psi_{-}(0)=\sqrt{
\frac{1-n_{z}}{2}}ie^{i\gamma}\left\vert 0\right\rangle + \sqrt{\frac{1+n_{z}
}{2}}\left\vert 1\right\rangle$ with eigenvalue $-ie^{-i\theta}$, where
$\gamma=\arctan(n_{x}/n_{y})$. Hence,
adopting $\Psi_{+}(0)$ as the time-zero state, we obtain the results
\bea P_{F}(\tau)=P_{I}(0)=\frac{1+n_{z}}{2}.
\eea
As an example, if $\epsilon =\pi /2$ and a strong pulse $\sigma _{x}$ kicks at
$t=0$ such that $\varpi \delta \rightarrow \pi/2$, 
in the limit $\delta \rightarrow 0$ one gets that $n_{z}\rightarrow 1$, $\Psi
_{+}(0)\rightarrow \left\vert 0\right\rangle $ and $U(\tau )\Psi
_{+}(0)\rightarrow \left\vert
1\right\rangle $, which is a \textit{complete} quantum transition from
$\left\vert 0\right\rangle $ to $\left\vert 1\right\rangle $, satisfying
$\left\vert
\left\langle I|\Psi (0)\right\rangle \right\vert =1$ and $\left\vert
\left\langle F|\Psi (0)\right\rangle \right\vert =0$.
%
In Fig. \ref{Fig2} we show that by carefully tuning the interaction
parameters, e.g. the delay time $\delta$, one can achieve such a
complete transition. We next demonstrate that a complete transition
can take place in general in an adiabatically evolving system.


\emph{Complete transitions: Adiabatic evolution.---} Consider a
time-dependent Hamiltonian $H(t)$. If it is varied sufficiently
slowly, the evolution of the system is adiabatic, and the system
occupies an (instantaneous) eigenstate of the Hamiltonian $H(t)$,
provided the time-zero state $\left\vert \Psi (0)\right\rangle $ is
an eigenstate of $H(0)$. If the state of the system at time $\tau$,
$\left\vert \Psi (\tau )\right\rangle$, is orthogonal to the
time-zero state, one can obtain a complete transition by setting
$\left\vert I\right\rangle =\left\vert \Psi (0)\right\rangle$ and
$\left\vert F\right\rangle =\left\vert \Psi (\tau )\right\rangle$,
both eigenstates of $W(\tau)$. As an example, consider the magnetic
Zeeman effect where a magnetic field splits the atomic (or
molecular) degenerate levels, characterized by the magnetic quantum
numbers $M$. The Hamiltonian is effectively given by
\bea H=B(\epsilon )J_{z}+T(\epsilon )J_{x},
\eea
%
where the second term $T(\epsilon )J_{x}$ is responsible to quantum
transitions between different values of $M(=-J,...J)$. The time
dependent modulation is controlled by the parameter $\epsilon
=\frac{\tau }{2}-t$, and we manipulate the magnetic field such that
$T(\epsilon )$ [$B(\epsilon )$] is an even [odd] function of
$\epsilon $, $T(\pm \frac{\tau }{2})=0$ and $B(0)>0$. We now choose
the initial state as $\left\vert I\right\rangle =$ $\left\vert
-J\right\rangle $, the lowest eigenstate of $H(0)$. If we control
the evolution of $H(t)$ adiabatically from time $0$ to $\tau $, the
state of the system at time $\tau $ becomes $\left\vert
F\right\rangle =$ $\left\vert
J\right\rangle $, which is the lowest eigenstate of $H(\tau )$. Since $%
|J\rangle $ and $|-J\rangle $ are orthogonal, the quantum transition (\ref%
{eq:exact}) is complete.
The results of Ref. \cite{Nori06} may be an example of this scenario when
$J=1/2$.


\emph{Superposition: Eigenstates in a three-level system.---}
Finally, we address the following conceptual question:
can one achieve an exact population transition (\ref{eq:est}) with a
superposition of the eigenstates of $W(\tau)$ as the time-zero state
of the system? We explain next the conditions for this transfer by
considering a three-level Hamiltonian \emph{\ }$H=\Omega (\left\vert
0\right\rangle \left\langle 1\right\vert +\left\vert 1\right\rangle
\left\langle
2\right\vert +h.c.)$. 
If one chooses the initial and final states to be $\left\vert
I\right\rangle =\left\vert 0\right\rangle $ and $\left\vert
F\right\rangle =\left\vert 2\right\rangle $, the exchange operator
$\mathcal{E}=\left\vert 0\right\rangle \left\langle 2\right\vert
+\left\vert 2\right\rangle \left\langle 0\right\vert +\left\vert
1\right\rangle \left\langle 1\right\vert $ commutes with $H$
\cite{comment}. One can easily obtain the eigenstates and
eigenvalues of $W(\tau)=\mathcal{E}U(\tau)$,
\begin{eqnarray}
\Psi_{a}(0) &=& - \frac{1}{\sqrt{2}}\left( \left\vert 0\right\rangle
-\left\vert 2\right\rangle \right) ;\,\,\ e^{i\phi_{a}}=-1 \nonumber
\label{eq:3LS} \\
\Psi _{b}(0) &=&\frac{1}{2}\left( \left\vert 0\right\rangle + \gamma_b
\left\vert 1\right\rangle +\left\vert 2\right\rangle \right) ;\,\,\ e^{i\phi
_{b}}=e^{-i\sqrt{2}\Omega \tau }  \nonumber \\
\Psi _{c}(0) &=&\frac{1}{2}\left( \left\vert 0\right\rangle +\gamma_c
\left\vert 1\right\rangle +\left\vert 2\right\rangle \right) ;\,\,\,e^{i\phi
_{c}}=e^{i\sqrt{2}\Omega \tau }.
\end{eqnarray}
$\gamma_{b,c}$ are functions of $\tau$ and $\Omega$ but their exact
form is not important for the discussion below. Examining
(\ref{eq:3LS}), it is obvious that there is no significant
transition if the time-zero state is any of these three eigenstates,
since $|\langle I|\Psi _{k}(0)\rangle |=|\langle F|\Psi
_{k}(0)\rangle |$ for $k=a,b,c $.
We show next that under some strict conditions, a superposition
state can yield a complete transition. Since the set $\left\{ \Psi
_{k}(0)\right\} $ is a complete orthogonal set, a general time-zero
state can be expanded as $\Psi(0)=\sum_{k=1}^{M}C_{k}\Psi_{k}(0)$,
not necessarily an eigenstate of $W(\tau )$,
\bea W(\tau )\Psi (0)=\sum_{k=1}^{M}C_{k}e^{i\phi _{k}}\Psi
_{k}(0).
\eea
An exact transition can still take place at a specific time $T$,
obeying the eigenvalue equation
\bea W(T)\Psi (0)=e^{i\phi (T)}\Psi (0). \label{eq:super} \eea
This equality is satisfied if
%
$C_{k}[\exp (i\phi _{k})-\exp (i\phi (T)]=0$.
%
Thus, for the contributing $k$ coefficients,
$C_{k}\neq 0$, we obtain a set of conditions $\phi(T)=\phi
_{k}+2\pi K_{k}$, where $K_{k}$ are arbitrary integers.
This is a very restrictive condition when there are many
nonzero coefficients. However, it may be still satisfied for particular
systems with special symmetry. We exemplify this within the
three-level model presented above [See Eq. (\ref{eq:3LS})]. Assuming
the following superposition state at time zero,
$\Psi(0)=C_{a}\Psi_{a}(0)+C_{b}\Psi_{b}(0)$, we obtain
\bea
W(\tau)\Psi (0)=-\left[ C_{a}\Psi_{a}(0)-e^{-i\sqrt{2}\Omega
\tau }C_{b}\Psi _{b}(0) \right],
\eea
which is generally not an eigenstate of $W(\tau)$. However,
Eq. (\ref{eq:super}) is satisfied at the specific time $\tau =\pi/(\sqrt{2}\Omega)$ leading to
\bea
\left\langle I|\Psi (0)\right\rangle &=&C_{b}/2 - C_{a}/\sqrt{2}  \nonumber \\
\left\langle F|\Psi (0)\right\rangle &=&C_{b}/2 + C_{a}/\sqrt{2},
\eea
manifesting a significant transition and a population transfer
\bea
P_{I}(0) &=&P_F(\pi /\sqrt{2}\Omega ) \\
&=& \frac{\left\vert C_{a}\right\vert
^{2}}{2}+\frac{\left\vert C_{b}\right\vert
^{2}}{4} - \frac{1}{2\sqrt{2}}\left( C_{a}^{\ast }C_{b}+C_{b}^{\ast
}C_{a}\right).  
\nonumber
\eea
The transition can be made complete, depending on the superposition
preparation coefficients $C_{b}$ and $C_{a}$.

\emph{Conclusion.---} We have proved the universality of exact
quantum transitions, demonstrating that for a given Hamiltonian and
a pair of states $|I\rangle$  and $|F\rangle$ in the associated
Hilbert space, there always exists a complete set of orthogonal
states that when employed as the time-zero state of the system, lead
to an exact population transition between the pair $|I\rangle$ and
$|F\rangle$. This universal proposition is a fundamental feature of
quantum dynamics and a promising building block in quantum control.
We have demonstrated the result analytically on the time-modulated
two-level-system model, showing that in some cases a complete
population transfer can be obtained. We have also shown that an
adiabatic evolution can lead to a complete transition. Finally, we
have analyzed exact population transitions in a superposition of
states.
Applications to specific control studies are the subject of future
work.

\noindent \emph{Acknowledgments.} LAW has been supported by the
Ikerbasque Foundation. DS acknowledges support from the University
of Toronto Start-up Fund, and PB was supported by NSERC.



\begin{thebibliography}{99}
\bibitem{comment} $\mathcal{E}$ does not commute in general with the
Hamiltonian. For example, if $\mathcal{E}=\left\vert 0\right\rangle
\left\langle 1\right\vert +\left\vert 1\right\rangle \left\langle
0\right\vert +\left\vert 2\right\rangle \left\langle 2\right\vert$,
then $[\mathcal{E},H]\neq 0$.
\end{thebibliography}
\end{document}